\documentclass[useAMS,usenatbib]{mn2e}
\usepackage{subfigure}
\usepackage[dvips]{graphicx}
\usepackage{times}
 \title[Radio continuum properties of young planetary nebulae]
{Radio continuum properties of young planetary nebulae}

\author[L. Cerrigone et al.]{L. Cerrigone,$^1$ \thanks{E-mail: lcerrigone@cfa.harvard.edu}
G. Umana$^2$, C. Trigilio$^2$,
 P. Leto$^3$, C. S. Buemi$^2$ and J. L. Hora$^1$ \\
   $^1$Harvard-Smithsonian Center for Astrophysics,
              60 Garden St, MS-65, Cambridge, MA, 02138, USA\\
    $^2$INAF-Osservatorio Astrofisico di Catania, via S. Sofia 78, 95123 Catania, Italy \\
   $^3$INAF-Istituto di Radioastronomia, C.P. 161, Noto, Italy \\}

\begin{document}

   \date{Received September 15, 1996; accepted March 16, 1997}

 \pagerange{\pageref{firstpage}--\pageref{lastpage}} \pubyear{2008}

\maketitle

\label{firstpage}

\begin{abstract}

We have selected a small sample of post-AGB stars in transition towards the
planetary nebula and present new Very Large Array multi-frequency high-angular resolution
radio observations of them. The multi-frequency
data are used to create and model the targets' radio continuum spectra,
proving that these stars started their evolution as very young
planetary nebulae. In the optically thin range, the slopes are compatible
with the expected spectral index (-0.1). Two targets (IRAS 18062+2410
and 17423-1755) seem to be optically thick even at high frequency, as
observed in a handful of other post-AGB stars in the literature, while a
third one (IRAS 20462+3416) shows a possible contribution from cold dust.
In IRAS 18062+2410, where we have three observations spanning a period of
four years, we detect an increase in its flux density, similar to that observed in CRL 618.

High-angular resolution imaging shows bipolar structures that may be due to circumstellar tori, although a different hypothesis
(i.e., jets) could also explain the observations. Further observations and
monitoring of these sources will enable us to test the current
evolutionary models of planetary nebulae.

\end{abstract}


\maketitle
%

\section{Introduction} 

The evolution of planetary nebulae remains a challenging topic in
astrophysics. Several studies have focused on the link between the
post-Asymptotic Giant Branch (AGB) and planetary nebula phases
to investigate the changes that occur in the evolutionary stage between these
object classes. Planetary nebulae (PN) have been observed over a wide
wavelength range, from X-ray to radio frequencies. Their complex
morphologies and the shaping mechanisms that produced them are still a
matter of debate.  Companion stars, jets from central stars, magnetic
fields, dust tori, and interacting winds are some of the possible shaping 
agents suggested as being responsible for various PN morphologies, 
and an overlap of their actions cannot be ruled out.

Hubble Space Telescope (HST) imaging has demonstrated that pre-PN show
non--spherical symmetry even at late spectral types, which implies that the
shaping process in these sources starts very early after the AGB
\citep{balick&frank}. Non-spherically symmetric structures have also been
observed in AGB stars that are part of a binary system \citep{karovska},
which could indicate that some shaping agents are already active in this
evolutionary phase. In general, the current theory of PN evolution is
based on the Interacting Stellar Wind (ISW) model \citep{kwok1978} and its
generalised version \citep{kahn}. However, this model does not account for
shaping because it assumes that an asymmetric distribution of matter is
already present when the wind interaction occurs. Other models take into
account the possible role of jets, and have been
successfully applied to some nebulae \citep{sahai98}. It has been
suggested that the interaction with a companion object, even a massive
planet, may provide the necessary asymmetry
\citep{soker06}. Also, large scale magnetic fields, influencing or
determining the shapes, might be sustained by a dynamo process
\citep{blackman}.

The birth of a PN is defined by the observation of an ionisation front,
which is itself a shaping agent and could heavily influence the morphology
established in earlier evolutionary phases, disrupting the molecular and
dust circumstellar shells. Near- and mid-IR images of PN, which trace the
molecular and warm dust emission, have been compared to optical line
images, which trace the ionised elements, and have shown the presence of
similar structures \citep{latter}. Therefore it seems that molecular and
ionised gas and dust grains can spatially coexist in these sources. This
makes the investigation of the spatial distribution and
physical properties of the ionised component in these envelopes even more compelling.

In this context, important information can be provided by observations of
very young PN, or pre-PN, where the physical processes associated with PN
formation are still occurring. To investigate the properties of these rare
objects in transition from the post-AGB to PN, we have selected a sample
of pre-PN and searched for radio emission from ionised shells. The targets
were selected from stars classified in the literature as hot post-AGB
candidates, showing strong far-IR excess and B spectral type features. We
detected radio emission in 10 sources in the selected sample
\citep{umana}. The detection of radio continuum emission is proof of the
presence of free electrons and therefore of an ionised shell.

For a better understanding of these sources, we have performed a
multi-frequency follow-up to build up radio continuum spectra and collect
information about the physical conditions in the nebulae. Several nebular
parameters can be determined by radio flux density measurements (for
example, electron density and ionised mass) but most calculations rely on
the assumption of an optically thin emission. Although this is presumably
the case for observations at frequencies higher than 5 GHz, only the
construction of the whole centimetre continuum spectrum can enable us to
distinguish between optically thick and thin regimes.  VLA A array
observations have also given us insight in the morphologies of the
envelopes.


\section{Observations and results}

\subsection{Low-angular resolution observations}

Multi-frequency observations were carried out at the Very Large Array,
managed by NRAO, in several runs, at 1.4, 4.8, 8.4, 14.9, and 22.4 GHz
(respectively 20, 6, 3.6, 2, and 1.3 cm), in D and C arrays, to
inspect the spectral energy distribution in the radio range. Table
\ref{tab1} lists the observation details. The 8.4 GHz observations were
performed in 2001 \citep{umana}. During the 2003 runs, strong interference
led to the failure of almost all of the 1.4 GHz observations and part of
the 14.9 GHz ones. To acquire information about the lower frequency range
of the spectrum of four sources (namely IRAS 18062+2410, 19336-0400,
19590-1249, 20462+3114), observations were performed in 2005 at both 1.4
and 8.4 GHz; the latter ones enabled us to check for any intrinsic
variation in the emitted flux. Although these latter runs were carried out
in a different array configuration (with a smaller synthesised beam), this
cannot cause missing flux, since all of our sources are point-like in both
D and C arrays at any frequency. The Largest Angular Scale
(LAS)\footnote{The LAS is the size of the largest structure that
an interferometer can map.} is much larger than the typical sizes of
these objects ($\le2''$) even at the shortest wavelength, being 60$''$ at
22.4 GHz (D and C arrays have the same LAS). The nominal beam
sizes in D array are 44$''$, 14$''$, 8.4$''$, 3.9$''$, and 2.8$''$, while
in C array they are 12.5$''$, 3.9$''$, 2.3$''$, 1.2$''$, and 0.9$''$,
respectively at 1.4, 4.8, 8.4, 14.9, and 22.4 GHz. One of the nebulae
detected in 2001 (18371-3159) could not be observed because of limited
observation time.

\begin{table*}
\centering
\caption{Summary of all the performed observation runs.}
\label{tab1}
\begin{tabular}{@{}lcccc}
\hline
             \multicolumn{5}{c}{Low-angular resolution} \\
\hline
Date &  VLA Conf. & Freq. (GHz) & Flux Calibr. & Targets \\
\hline
2003 &            &             &              &          \\
Feb 15 & D & 1.4, 4.8, 8.4, 14.9, 22.4 & 3C286 & 18062+2410 \\
             &	 &			   &       & 17423-1755 \\
             &   &                         &       & 17460-3114     \\
             &   &                         &       & 20462+3416 \\  
\noalign{\smallskip}    
Feb 16 & D &  1.4, 4.8, 8.4, 14.9, 22.4 & 3C286 & 20462+3416 \\
             &   &                          &       & 19336-0400 \\
             &   &                          &       & 18442-1144 \\
             &   &                          &       & 19590-1249 \\
\noalign{\smallskip}
Mar 8 & D & 1.4, 4.8, 8.4, 14.9, 22.4  & 3C286 & 17381-1616 \\
             &   &                          &       & 17460-3114 \\
            &   &                           &       & 18062+2410 \\
\noalign{\smallskip}
Mar 21 & D & 1.4, 4.8, 8.4, 14.9, 22.4 & 3C48 &  06556-1623 \\
\hline
2004  &    &                           &      &              \\
Apr 18 & C & 8.4 & 3C48 & 22023+5249 \\
\hline
2005   &   &     &      &           \\
Jul 25 & C & 1.4 & 3C48 & 18062+2410 \\
             &   &     &      & 19336-0400 \\
             &   &     &      & 19590-1249 \\
\noalign{\smallskip}
Jul 26 & C & 1.4, 8.4 & 3C48 &  18062+2410\\
             &   &          &      & 19336-0400 \\
             &   &          &      & 19590-1249 \\
             &   &          &      & 20462+3416 \\
\hline
\multicolumn{5}{c}{High-angular resolution} \\
\hline
2003   &   &     &     &           \\
Jul 10 & A & 8.4 & 3C48 & 18062+2410 \\
             &   &     &      & 18442-1144 \\
             &   &     &      & 19590-1249 \\
             &   &     &      & 19336-0400  \\
\noalign{\smallskip}
Aug 15 & A & 8.4 & 3C48 & 17381-1616 \\
\hline
2004      &   &     &     &           \\
Oct 29 & A & 8.4 & 3C286 & 22023+5249 \\
\hline
\end{tabular}
\end{table*}

A typical observing scan consisted of two 15-minute pointings on target,
preceded and followed by 1-minute scans on the phase calibrator. For a
better phase calibration, 22.4 GHz observations were carried out pointing
at the target for four 7-minute scans alternating with 1-minute scans on
the phase calibrator.

The data were reduced with the \textbf{A}stronomical \textbf{I}mage
\textbf{P}rocessing \textbf{S}ystem (AIPS), according to the recommended
reduction process: the data set was FILLMed into AIPS with average opacity
correction and nominal sensitivities (\textit{doweight}=1), edited to flag
interference or any other bad points and CALIBrated along with point
weights (\textit{docalib}=2). Maps were obtained using the task IMAGR with
natural weights and CLEANed down to the theoretical rms noise whenever
possible, performing some hundreds of iterations of the CLEAN algorithm.  
The flux density for each source in every map was estimated by fitting a
Gaussian to the unresolved source (task JMFIT), and the rms noise
was calculated in an area much larger than the synthesised beam ($>$100
beam), without evident sources in it (task IMEAN).

For IRAS 17423-1755 and 06556+1623, a second nearby (within 40$''$) source
was detected and large wavelength data cannot be considered indicative of
their flux densities because of confusion problems. Tables~\ref{tab1} and
\ref{tab2} summarize the observation runs and the flux density
measurements obtained. The reported error has been calculated as $\sigma
=\sqrt{rms^2+(\sigma_{cal} S_{\nu})^2}$, where $\sigma_{cal}$ is a 5\%
absolute calibration error.



\begin{table*}
\centering
\caption{Flux densities and their errors (including a 5\% absolute flux
density calibration error) from our D and C array observation runs are
shown for each source and frequency. The asterisks indicate the sources
chosen for high-angular resolution imaging. The spectral indexes in the
optically thin range and the rms in the high angular resolution maps are
also shown.}
\label{tab2}
\begin{tabular}{@{}l|cccccccc}
\hline
IRAS ID 	& \multicolumn{5}{c}{Flux density (mJy)} & Spectral index & A array rms (mJy) & A array beam size ($''$) \\
	& 1.4 Ghz	        & 4.8 Ghz		& 8.4 GHz		& 14.9 GHz		& 22.4 GHz	&			& 8.4 GHz	 & 8.4 GHz	\\
\hline
06556-1623 	& 2.0$\pm$0.1$^{1}$ 	& 0.60$\pm$0.04 	& 0.55$\pm$0.03		& 0.43$\pm$0.09		& 0.36$\pm$0.06 & -0.29$\pm$0.20	& --	 & --	\\
17381-1616*	& 1.1$\pm$0.1 		& 1.61$\pm$0.09 	& 1.42$\pm$0.09		& 1.3$\pm$0.1	 	& 1.19$\pm$0.08 & -0.20$\pm$0.11	& 0.038	 & 0.26	\\
17423-1755 	& --$^{2}$ 	        & 1.74$\pm$0.06$^{1}$ 	& 0.26$\pm$0.03 	& --$^{2}$		& 0.75$\pm$0.08 & +1.08$^5$		& --	 & --	\\
17460-3114*	& --$^{2}$	        & 1.6$\pm$0.1 		& 1.29$\pm$0.08 	& 1.7$\pm$0.1 		& 1.24$\pm$0.08 & -0.10$\pm$0.12	& --	 & --	\\
18062+2410* 	& 0.26$\pm$0.04$^{3}$ 	& 1.27$\pm$0.07 	& 1.46$\pm$0.09$^4$ 	& --$^{2}$		& 1.8$\pm$0.1 	& +0.22$\pm$0.10	& 0.045	 & 0.22	\\
18442-1144* 	& --$^{2}$	        & 17.8$\pm$0.9 	        & 19.2$\pm$0.9		& 14.2$\pm$0.7 		& 13.5$\pm$0.7 	& -0.22$\pm$0.09	& 0.038	 & 0.28	\\
19336-0400* 	& 8.0$\pm$0.4$^{3}$     & 10.4 $\pm$0.5 	& 9.7$\pm$0.5	 	& 8.0$\pm$0.4 		& 7.0$\pm$0.3 	& -0.26$\pm$0.09	& 0.034	 & 0.27	\\
19590-1249* 	& 3.2$\pm$0.2$^{3}$	& 3.1$\pm$0.2 		& 2.8$\pm$0.1		& 2.3$\pm$0.1 		& 2.1$\pm$0.1 	& -0.26$\pm$0.10	& 0.028	 & 0.32	\\
20462+3416	& 0.48$\pm$0.06$^{3}$ 	& 0.60$\pm$0.05 	& 0.42$\pm$0.05 	& 0.50$\pm$0.09 	& 0.83$\pm$0.07 & -0.02$\pm$0.17$^6$	& --	 & --	\\
22023+5249	& --			& --			& 2.0$\pm$0.1		& --			& --		& --			& 0.027	 & 0.29	\\
\hline
\end{tabular}
\flushleft
$^{1}$ More than one source in the beam: the flux density must be regarded as an upper limit. \\
$^{2}$ The data were not recovered. \\
$^{3}$ The measurement was performed in 2005.\\
$^{4}$ In 2005 the flux density was 2.1$\pm$0.1 mJy. \\
$^5$ This value is calculated with two data points only. \\
$^6$ The 22.4 GHz point was not included in the fitting procedure.
\end{table*}

\subsection{High-angular resolution observations}

To inspect the morphology of our targets, we carried out
observations at 8.4 GHz in A array at the VLA for those sources that were
brighter than 1~mJy at this frequency in our detection experiment reported
in \citet{umana}: IRAS 17381-1616, IRAS 18062+2410, IRAS 18442-1144, IRAS
19336-0400, and IRAS 19590-1249. The 1 mJy limit was set to select nebulae
with surface brightnesses high enough to be imaged with the A array. The
angular resolution of these observations was $\sim$0.2$''$. Two runs were
performed in 2003: the first on July 10 (UT 07:04:50; 10:29:30) and the
second one on August 15 (UT 05:12:30; 06:37:50). The integration time per
object was about 1.5 hours. The reduction process followed the standard
format, as reported in the previous section. In the task IMAGR, the pixel
size was set to 0.05$''$, so that the convolution beam is
at least three pixels wide;  Brigg's weighting of the data was
used to find a good compromise between sensitivity and resolution
(\textit{robust}=~0) for all the resolved sources except 19590-1249, as
natural weights were necessary in order not to miss its faint emission.  
For 17381-1616 and 18062+2410, we set respectively \textit{robust}=~-2 and
\textit{robust}=~-4 (nearly pure uniform weighting) to increase the
resolution. The number of CLEAN iterations was set to 1000,
\textit{minpatch} 127, and \textit{gain} 0.1. The absolute calibrator was
3C48, its flux density set to 3.15~Jy. On October 29 2004 we also
managed to perform a snapshot (15 minutes on source) in A array of IRAS
22023+5249, which we had detected as a radio source in a previous
observation run in C array and had measured as 2.0$\pm$0.1~mJy (including
absolute flux density calibration error).

\begin{figure*}
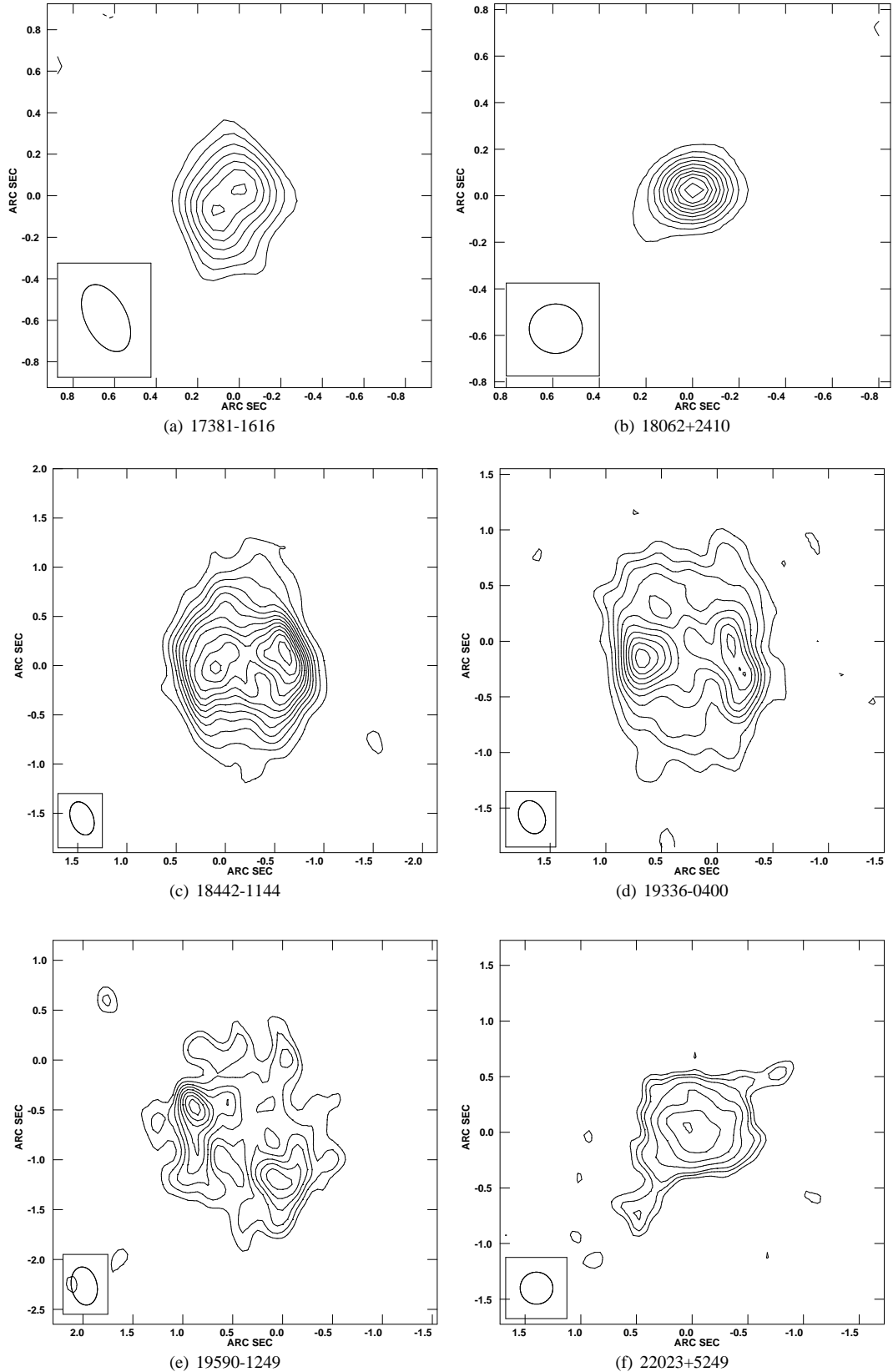

\centering
\subfigure[17381-1616]{\includegraphics[width=7.5cm,angle=270,scale=0.9]{fig1a.eps}}
\subfigure[18062+2410]{\includegraphics[width=7.5cm,angle=270,scale=0.9]{fig1b.eps}}
\smallskip
\subfigure[18442-1144]{\includegraphics[width=7.5cm,angle=270,scale=0.9]{fig1c.eps}}
\subfigure[19336-0400]{\includegraphics[width=7.5cm,angle=270,scale=0.9]{fig1d.eps}}
\smallskip
\subfigure[19590-1249]{\includegraphics[width=7.5cm,angle=270,scale=0.9]{fig1e.eps}}
\subfigure[22023+5249]{\includegraphics[width=7.5cm,angle=270,scale=0.9]{fig1f.eps}}

\caption{Radio maps at 8.4 GHz of the targets observed in high angular
resolution. The flux density levels, in units of $\mu$Jy/beam, are: in
17381 (-3, 3, 4, 5, 6, 7, 8, 9)$\sigma$; in 18062 (-3, 3,
6, 9, 12, 15, 18, 21, 24, 27, 30)$\sigma$; in 18442 (-3, 3, 6, 9,
12, 15, 18, 21, 24, 27, 30, 33, 36)$\sigma$; in 19336 (-3, 3, 5, 7, 9,
11, 13, 15, 17, 19, 21)$\sigma$; in 19590 (-3, 3, 4, 5, 6, 7, 8, 9,
10)$\sigma$; in 22023 (-3, 3, 4, 5, 7, 9, 11, 13)$\sigma$; where $\sigma$
indicates the rms calculated in each map and listed in
Table~\ref{tab2}. The maps do not show any negative level,
because no such levels are present in the mapped areas.}

\label{fig:radiomaps}
\end{figure*}

All the selected targets showed clear bipolar structures except IRAS
17381-1616 and IRAS 18062+2410, which were barely (IRAS 17381-1616) or not
(IRAS 18062+2410) resolved. Figure~\ref{fig:radiomaps} shows the
structures found in all the radio envelopes as observed with the A array.

\section{Radio properties}

\subsection{Modelling}

We have modelled the radio continuum spectra assuming that the central
star is surrounded by a shell of ionised gas with density radially
decreasing as $r^{-2}$.  We have calculated a 3D distribution of mass in a
spherical shell with outer radius R$_{out}$, inner radius
R$_{in}=\eta$R$_{out}$ (with $0<\eta<1$), and density at the inner radius
$\rho_{in}$. The calculation of the average density in the shell links this parameter to the density at the inner radius: $<\rho>\approx\eta^{2}(1-\eta)\rho_{in}$. The electron temperature was set
to 10$^4$~K for all nebulae. Details of the radio model can be found in
\citet{sao}.

Only a few estimates of the distances to the stars in our sample are
available. Where available in the literature, we have used the most common
estimates of this parameter. For IRAS 18442-1244 and 19336-0400, since
their sizes and flux densities are comparable to what detected in IRAS
19590-1249, we assume the same distance as for this source. For the
remaining targets we have adopted a standard value of 1~kpc as an estimate
of the order of magnitude of their distances. Our radio model code
requires as an input the distance to the star as well as the
density at the inner radius of the shell, outer radius of the ionised
shell, inner to outer radius ratio, and electron temperature. For the
outer radii we have assumed as upper limits the sizes estimated in
\citet{umana}, which were calculated from the VLA theoretical angular
resolution.
For those objects that were observed with the VLA in A array (IRAS
17381-1616, 18062+2410, 18442-1144, 19336-0400, and
19590-1249)\footnote{We do not have multi-frequency observations of IRAS
22023+5249, since we had not radio detected this target at the time of our
multi-frequency run.}, we used our estimate of their radii, based on their
3$\sigma$ size. For IRAS 18062+2410 it was necessary to use a different
geometry, since with a shell structure we could not model a steep
enough increase of flux density in the optically thick range. Therefore we
used a sphere instead of a spherical shell.

The results of our radio models are reported in Table~\ref{tab3}. The
density and ionised mass values are in agreement with those expected for
young planetary nebulae. For example, in NGC 7027, a high excitation young
PN, values of 10$^4$ cm$^{-3}$ and 0.005 M$_\odot$ have been measured
\citep{bains}, although, even in this very well studied PN, these results
suffer from the distance uncertainty. Such distance independent parameters
as emission measure and brightness temperature have been previously
calculated for our targets \citep{umana} and found to match the expected
values for young PN.

\begin{table*}
\centering

\caption{Model parameters for the radio shells. For 18062 a solid sphere
geometrical model was used. IRAS 18062+2410 was modelled as an
emitting sphere, not a shell.}

\label{tab3}
\begin{tabular}{@{}lcccccc}
\hline 
Target & Distance & R$_{out}$ & R$_{in}/$R$_{out}$ & $\rho_{in} $                 & $<\rho>$             & M$_{ion}$ \\
       & kpc      &   $''$    &                    &    $10^4$ cm$^{-3}$          &  $10^3$ cm$^{-3}$ & $10^{-4}$ M$_\odot$ \\
\hline 
06556-1623  & 1        &  0.5      &     0.15           &   15                         & 3                 & 0.1 \\
17381-1616  & 1        & 0.35      &  0.195             &      30                      & 9                 & 0.14 \\
17460-3114  & 1        & 1.1      & 0.049               &     40                       & 0.9               & 0.4\\
18062+2410  & 6.4~$^1$ & 0.061      & --                 &    25                        & 8                & 1.7 \\
18442-1144  & 4        & 1        & 0.225               &    8.8                       & 3.5               & 77 \\
19336-0400  & 4        & 1        & 0.2                 &    8                         & 2.6               & 57 \\
19590-1249  & 4~$^2$   & 1.2      & 0.28                &    2                         & 1.1               & 40 \\
20462+3416  & 3~$^3$   & 1.1      & 0.11                &    4                         & 0.04              & 5 \\
\hline
\end{tabular}
\flushleft
$^1$~\citet{bogdanov} \\
$^2$~\citet{conlon} \\
$^3$~\citet{arkhipova}
\end{table*}

We notice that in some cases the data points, especially at higher
frequency, seem to show a steeper than expected slope in the optically
thin range ($S_{\nu}\sim\nu^{-0.1}$). Besides our modelling, we estimated
the empirical spectral indexes by a linear fitting (in the
\textit{log-log} plane) of the data points that correspond to the
optically thin regime (typically 4.8--22.4 GHz data). The results of the
fitting are summarized in Table~\ref{tab2}.

Because at high frequency (14.9 and 22.4 GHz) a lack of short baselines
may cause a lower sensitivity to flux from extended structures (therefore
a lower total flux density), we checked whether tapering our data (i.e.,
assigning larger weights to short baselines) could lead to larger flux
densities, thus making the slopes in the spectra less steep. We did not
find any significant difference in the results from tapered and
non-tapered data. The empirical spectral indexes of three of the targets
(IRAS 06556-1623, 17381-1616, and 17460-3114) are close to the expected
value of -0.1, and out of the expected range for three others (IRAS
18442-1144, 19336-0400, and 19590-1249). Although we cannot rule out that
the observed steeper slopes might be real and further observations should
be carried out, if we consider that the reported errors are based on a
statistics on 4 points only, the deviation from -0.1 is not statistically
significant. Therefore we conclude that, within the errors, the spectral
indexes are compatible with their expected value.

\begin{figure*}
\centering
\subfigure{\includegraphics[width=7.5cm]{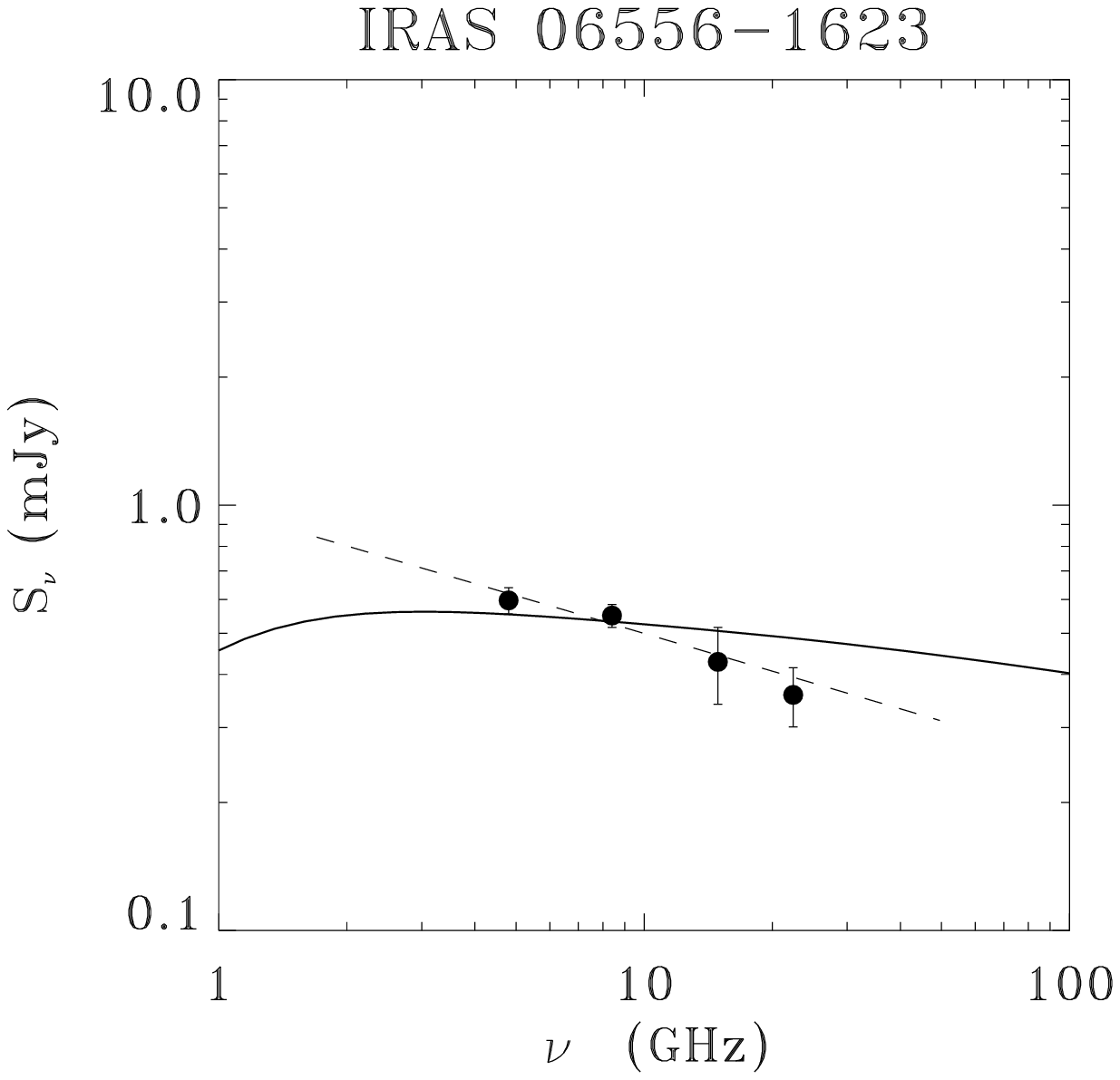}} 
\subfigure{\includegraphics[width=7.5cm]{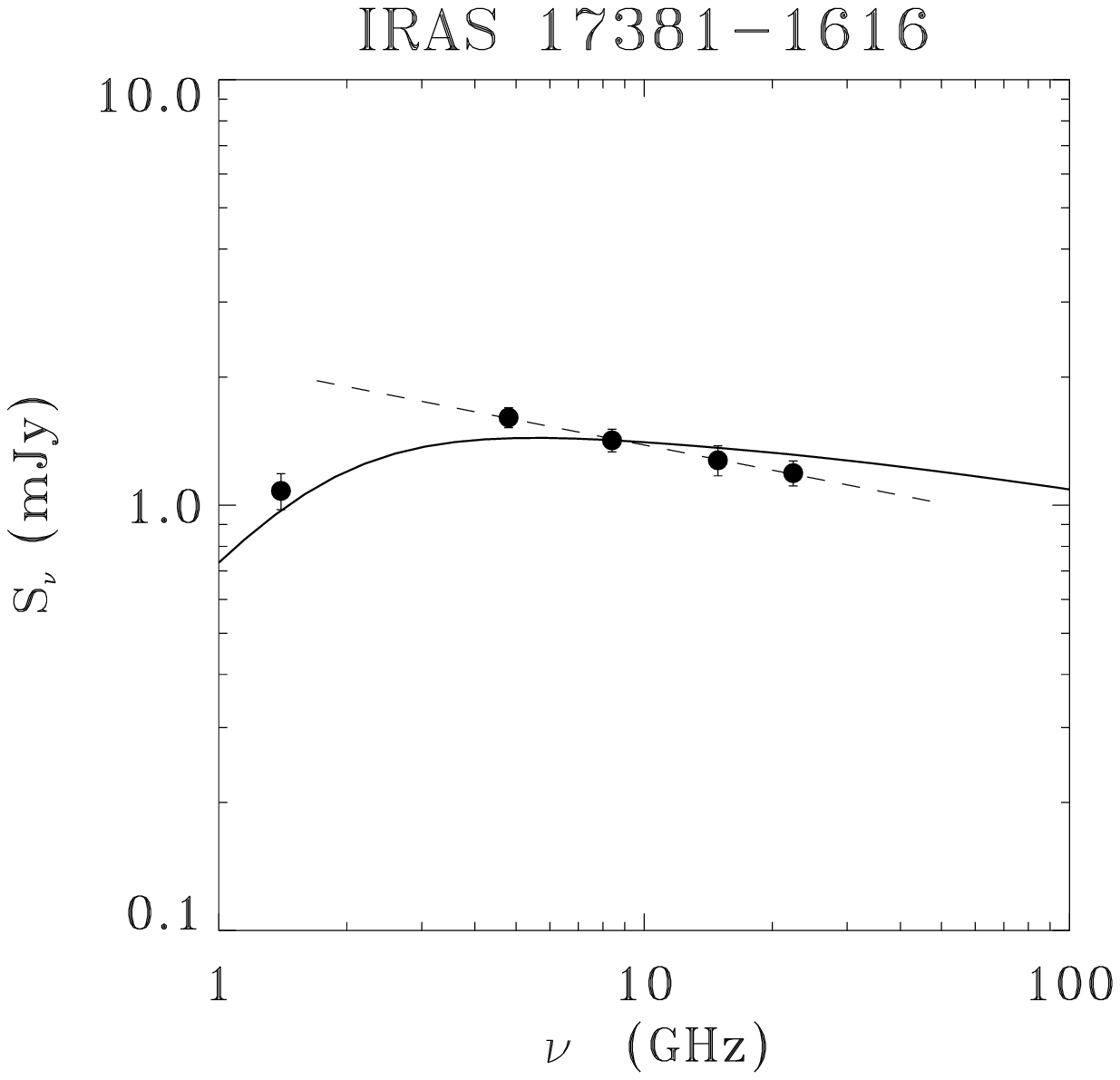}}
\smallskip
\subfigure{\includegraphics[width=7.5cm]{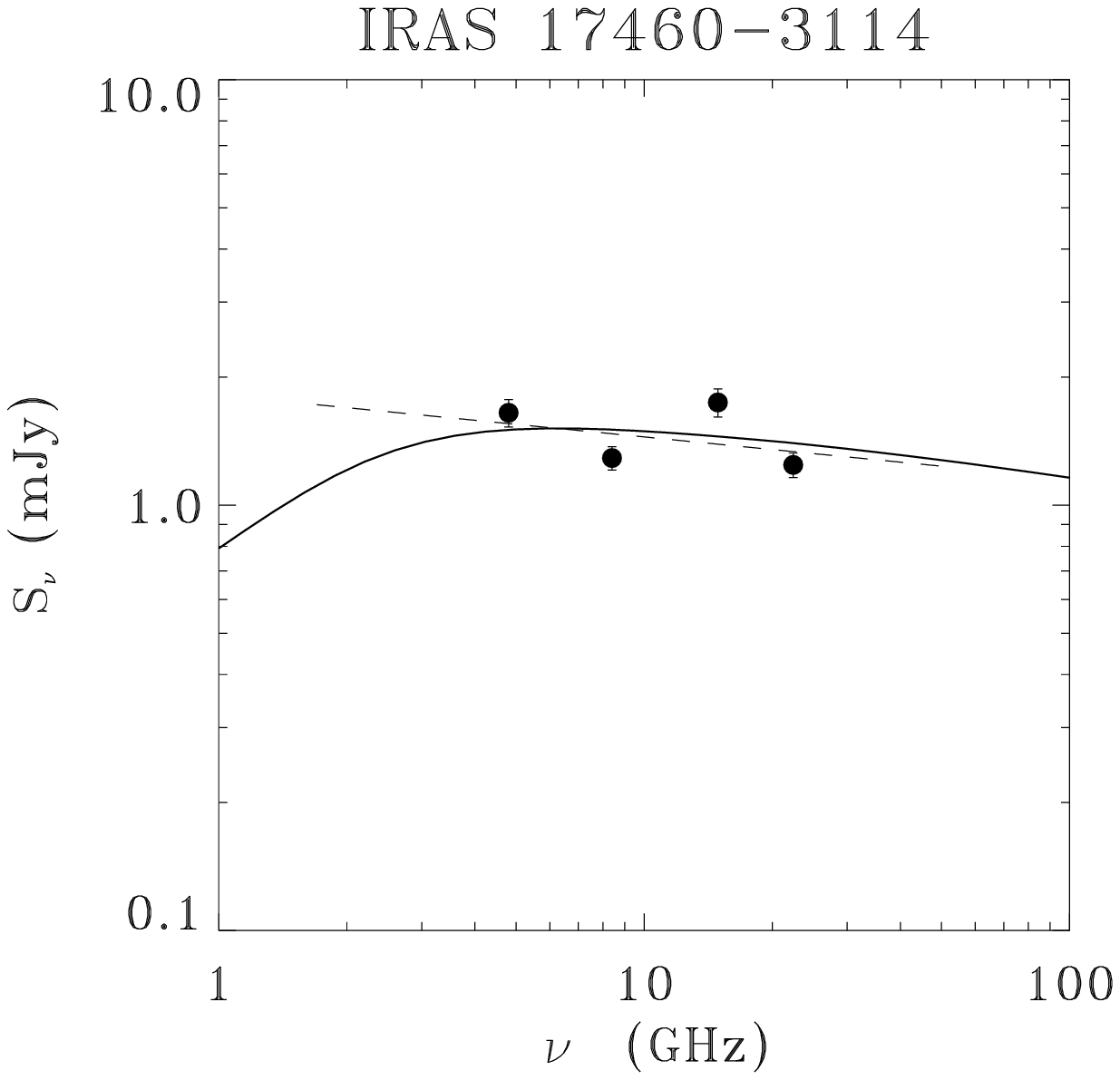}}
\subfigure{\includegraphics[width=7.5cm]{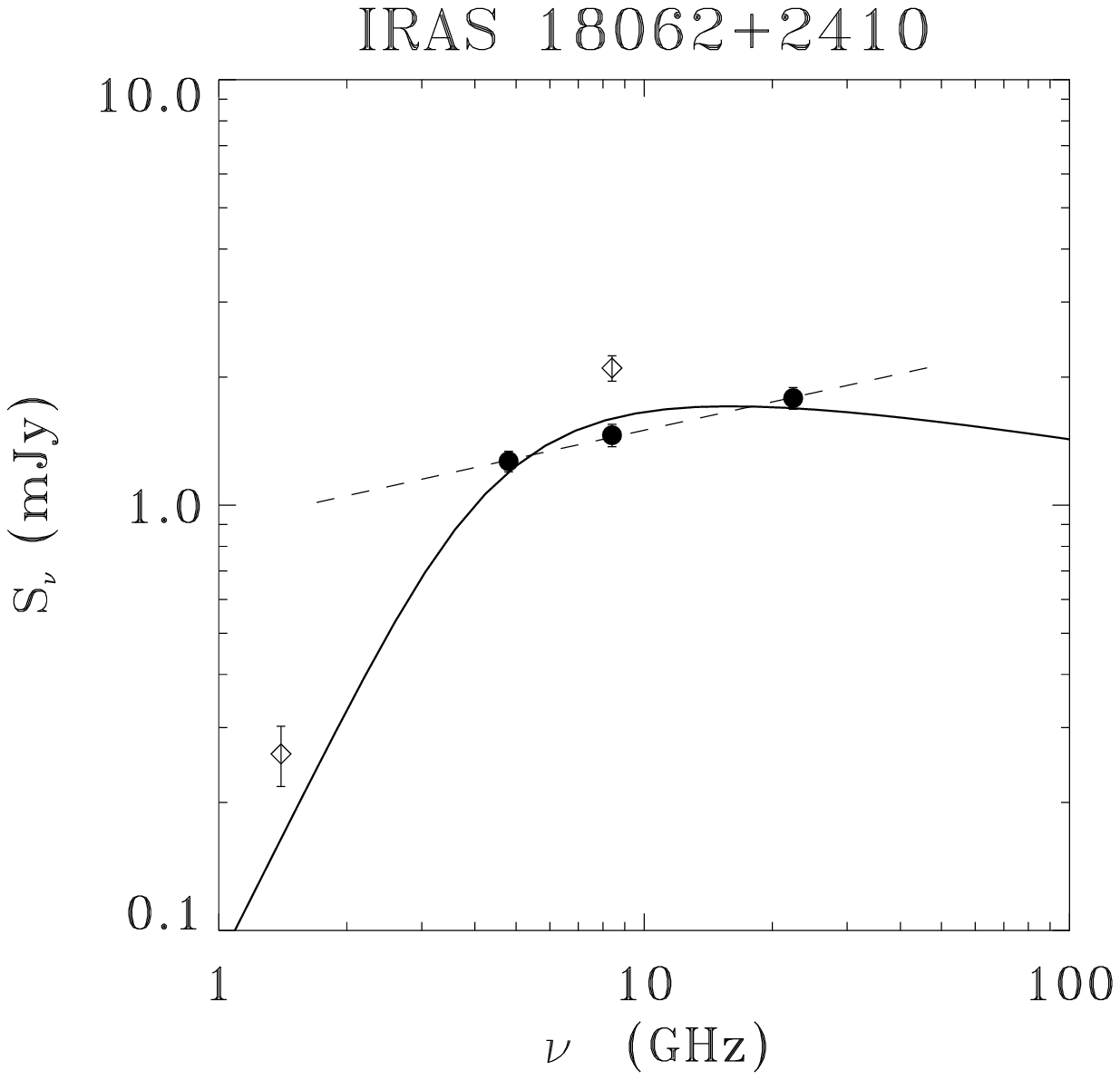}} 
\smallskip
\subfigure{\includegraphics[width=7.5cm]{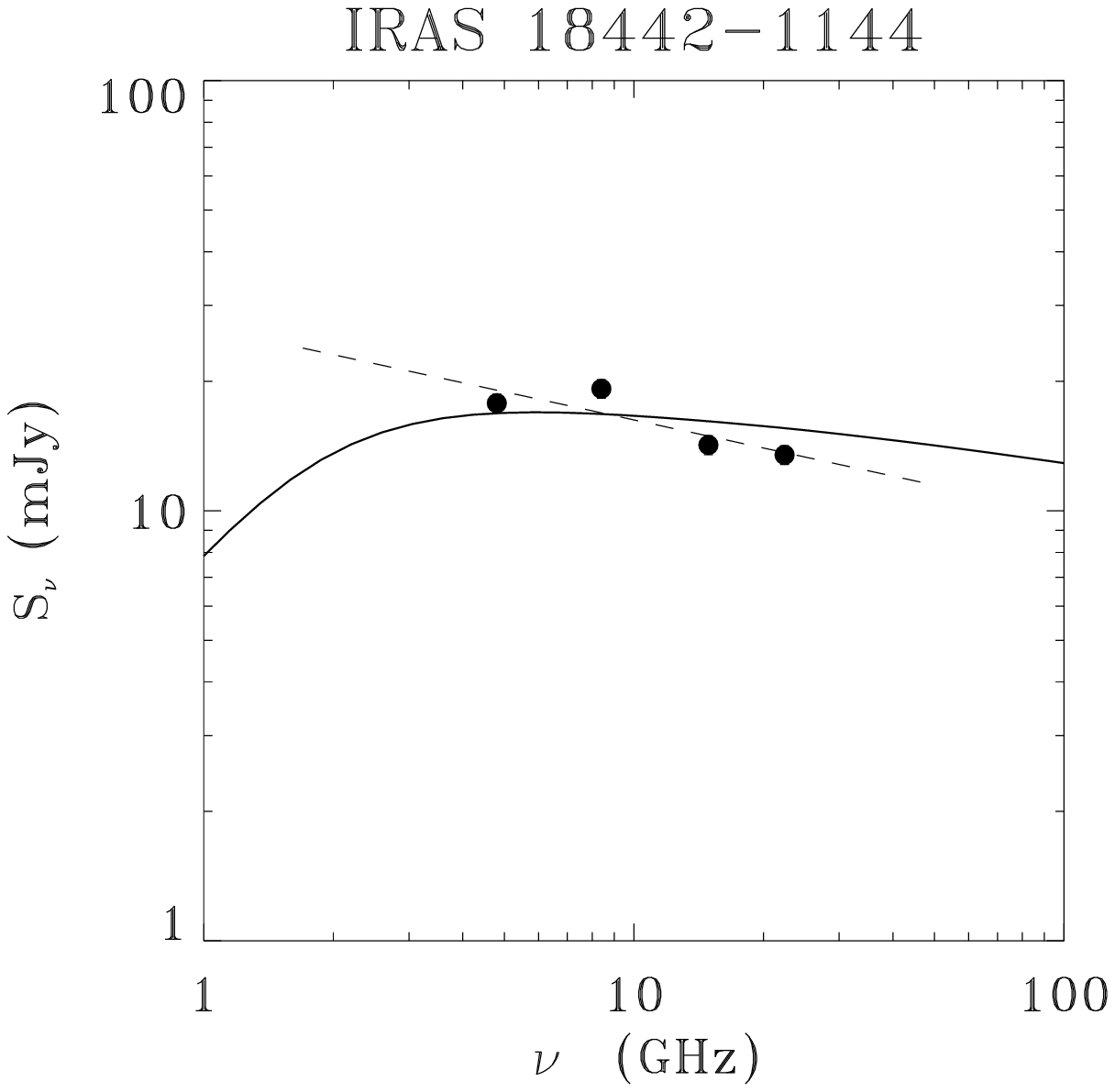}}
\subfigure{\includegraphics[width=7.5cm]{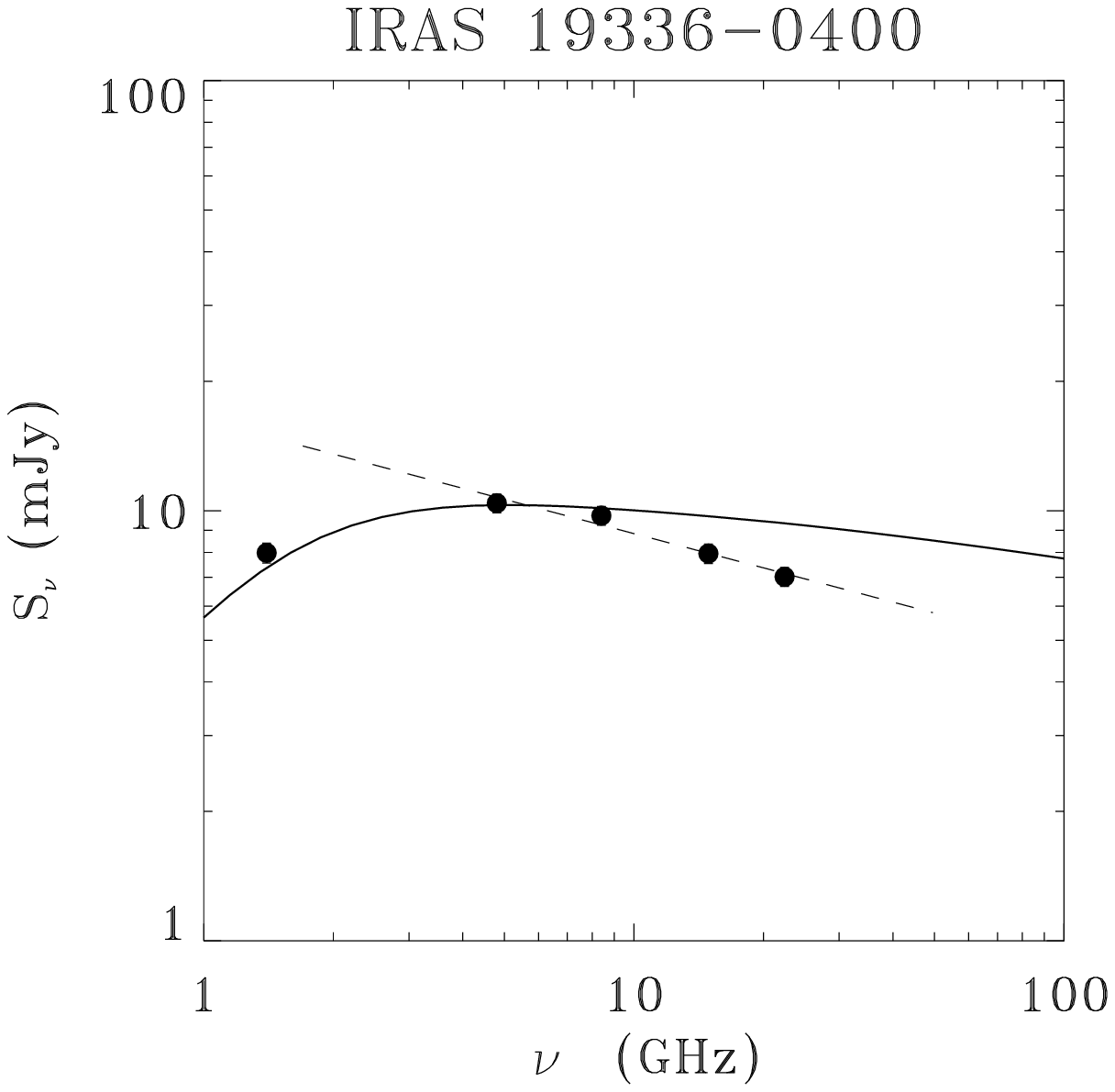}}
\smallskip
\subfigure{\includegraphics[width=7.5cm]{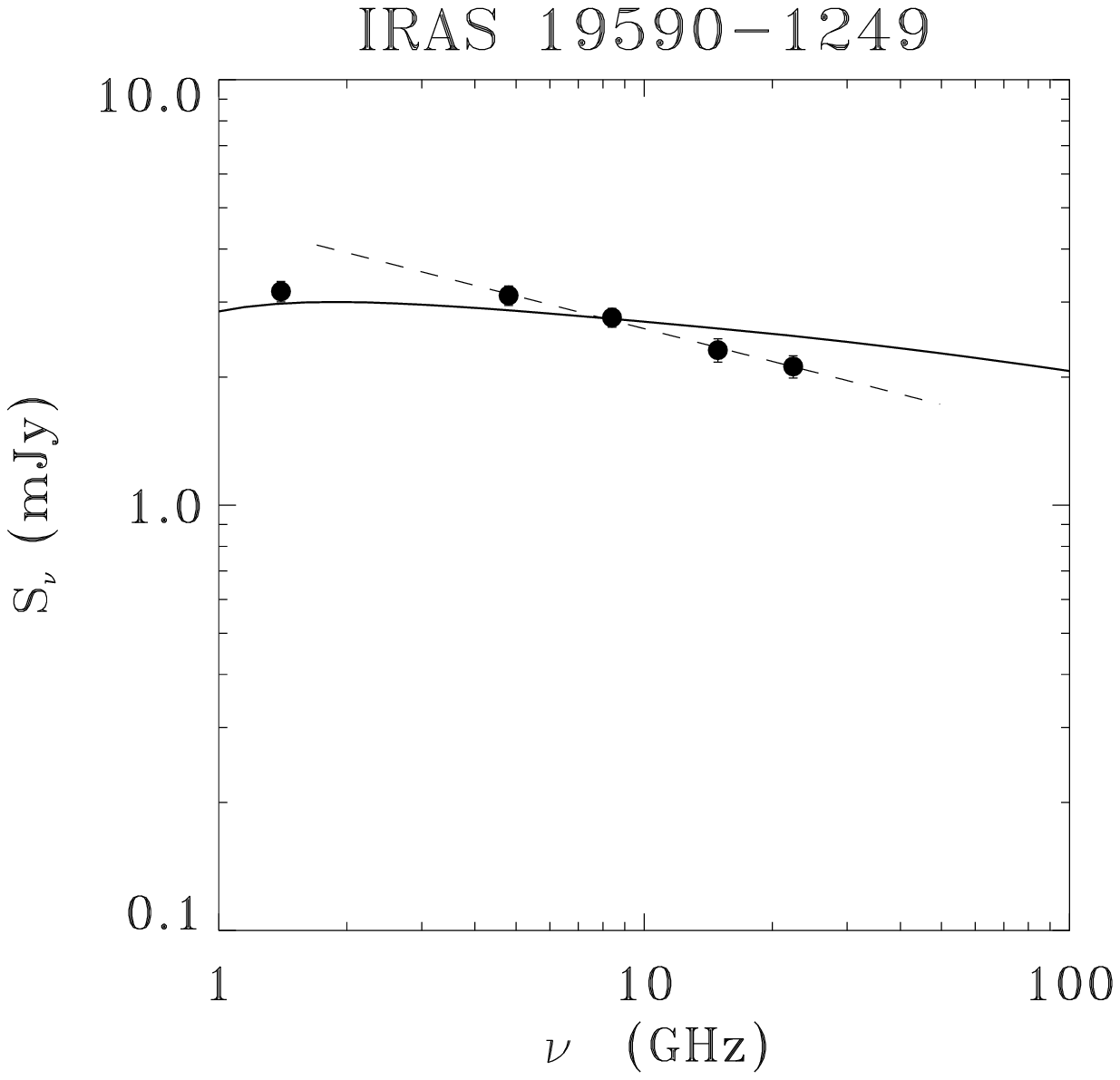}}
\subfigure{\includegraphics[width=7.5cm]{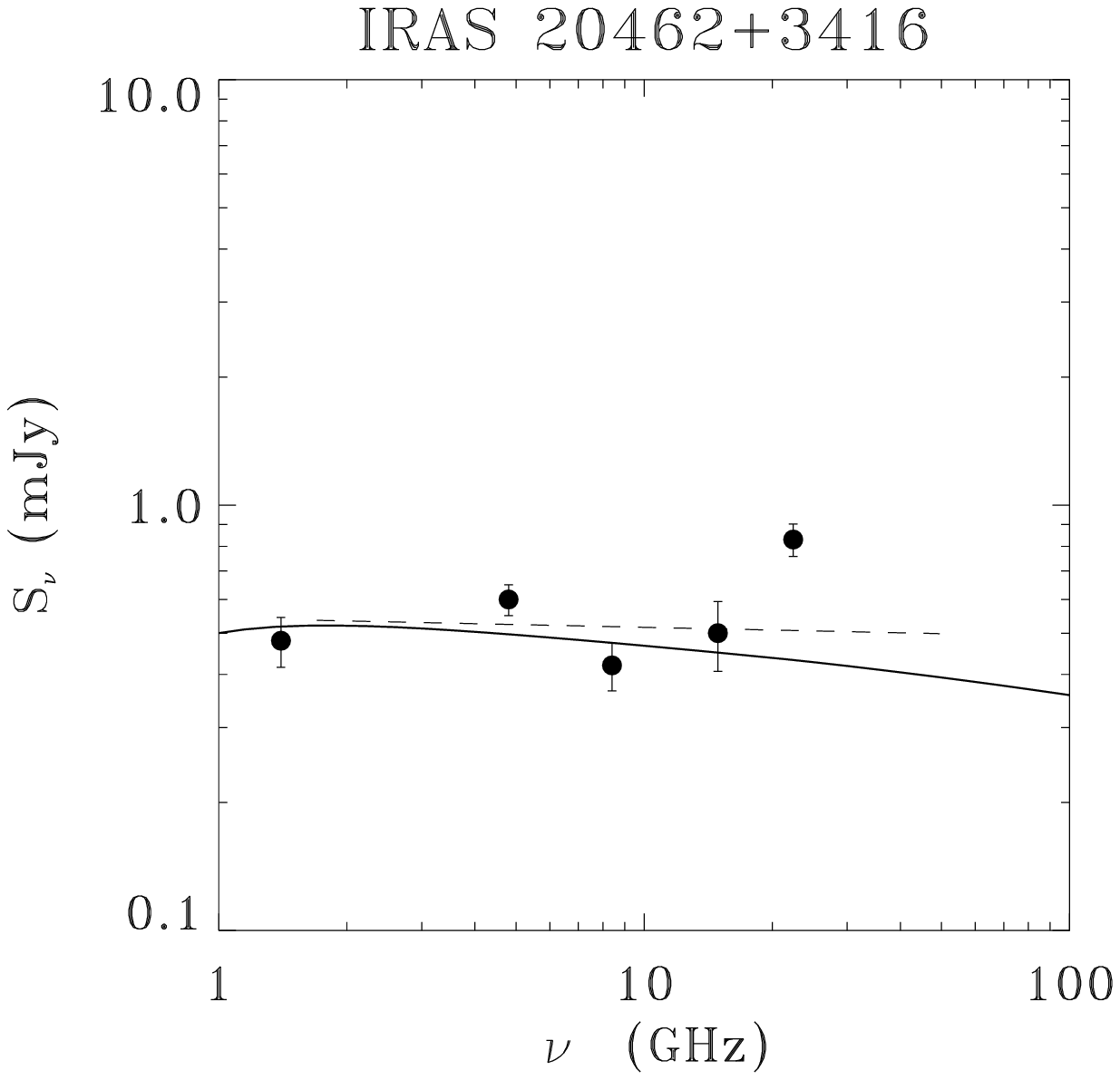}}

\caption[Radio continuum spectra]{Plots of the multi-frequency data
obtained at the VLA. For 18062+2410, 19336-0400, 19590-1249, and
20462+3416, 2005 1.4 GHz data points have also been used. In 18062 they
have been over-plotted as diamonds. The dashed lines are the result of the
linear (in the \textit{log-log} plane) fitting of the optically thin range
points. The plots therefore show the difference between the modelled
(solid line) and empirical (dashed line) dependence on frequency.  The
error bars are in some cases within the size of the data points.}

\label{fig:continuumplots}

\end{figure*}

Figure~\ref{fig:continuumplots} shows the data points obtained in our VLA
runs, the model spectra, and the linear fitting. We have not included
IRAS~17423-1755 in our modelling, because only two data points are
available. The flux density values obtained for this target would indicate
an optically thick spectrum up to 22 GHz.

Another particular result is the spectrum of IRAS~20462+3416, which is
well matched by the model in the 1--15 GHz range, while the 22.4 GHz point
is higher than the others.  Since the measured radio flux densities are
all below 1~mJy, even a small contribution from cold dust may be not 
negligible and so explain the 22.4 GHz measurement. Observations in the
mm and sub-mm ranges would assess whether such a cold dust component is present
or not.

\subsection{IRAS 18062+2410}

\begin{figure}
\centering
\includegraphics[width=7.5cm]{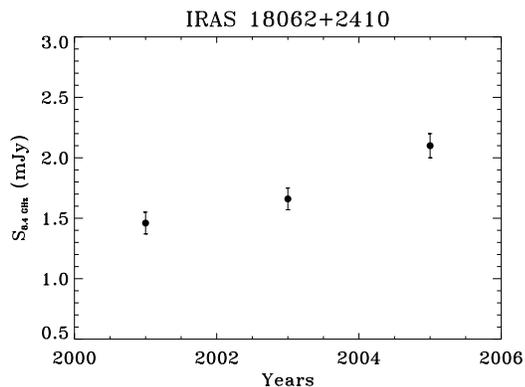}

\caption{Increase of the 8.4 GHz flux density in IRAS 18062+2410 during
years 2001--2005.}

\label{fig:18062incr}
\end{figure}

In our modelling we have used data from 2001, 2003, and 2005, since the
sources that were re-observed in 2005 (see above) did not show any flux
density variation with respect to the 2001 8.4 GHz results within observational
errors. The only exception is IRAS 18062+2410, whose flux density varied
by about +44\%, increasing from 1.46$\pm$0.09 in 2001 to 2.1$\pm$0.1~mJy
in 2005, while its secondary calibrator, 1753+288, varied from
0.82$\pm$0.03~Jy in 2001 to 1.41$\pm$0.04~Jy in 2005 (+75\%).
Unfortunately it was not possible to use the same primary calibrator in
the two runs (see Table~\ref{tab1}), which introduces a different density
scale for the two measurements, although the error should be within a few
percent (3-5\%). The flux density variation in the secondary calibrator
might indicate issues with the flux density calibration. Since other
targets were also observed in the same run and no change was measured in
their flux densities within the observational uncertainties, we conclude
that the variation observed in 1753+288 is real, as possible for a
secondary calibrator, but it does not account for the variation observed
in the target object.  When we combine all of our observations of this
target, including our 2003 measurement (A array data-set) of
1.66$\pm$0.09~mJy, we can conclude that the radio flux density of IRAS
18062+2410 has increased during the period we have monitored it (see
Figure~\ref{fig:18062incr}).

A similar uncommon increase of the radio flux density was detected in CRL
618 from 1975 to 1991 \citep{knapp}. If, as observed in CRL 618, we assume
a linear increase, we can fit the three points in
Figure~\ref{fig:18062incr} and extrapolate that the ionisation in IRAS
18062+2410 started around 1991, confirming the rapid evolution previously
observed in this object \citep{parthasarathy, arkhipova}.

The spherical model used in modelling this object allowed us to match the
2003 data, using the 2005 1.4 GHz value as an upper limit. We determine
that the shell in this target is optically thick beyond 5 GHz. Similar
behaviour has been observed in CRL 618 also \citep{knapp}. Further
observations, in both the radio and at other wavelengths, will help us to
determine whether these objects share other common features.

\subsection{Radio morphology}

Figure~\ref{fig:radiomaps} shows the structures observed in the resolved
targets.  All except IRAS 18062+2410 show bipolar radio morphology, with
clearly resolved peaks. IRAS 17381-1616 shows a hint of bipolarity,
although the angular resolution is not adequate to show its morphology.  
\citet{aaquist91} also observed a sample of young PN with high resolution
in the radio and found that most of their targets had a bipolar structure.
Such radio morphologies, to be set in the standard ISW scenario, require a
strong density gradient in the nebula, with higher density regions
perpendicular to the main axis of the nebula \citep{aaquist96}.  
Following this interpretation, the two bright peaks of emission are due to
an opacity effect, with the optically thin radio flux tracing higher
density regions. The other possibility is that we are actually observing
the very early onset of outflows/jets piercing the circumstellar envelope,
as evident in many HST images of pre-PN \citep{sahai98} and also pointed
out by high-resolution CO observations of the molecular component of the
envelope \citep{huggins}. However, it is difficult to interpret the
morphology observed in IRAS~22023+5349 in the framework of the standard
ISW model, even invoking a strong density gradient in the nebula.  For
this object, a jet would be a more likely source of the observed
morphology.

\section{Summary}

We have presented radio continuum observations of a small sample of
pre-PN. The observations were carried out to inspect both the morphology
and the physical conditions of the ionised component in these nebulae.
Compact planetary nebulae are, in principle, only one possible explanation
for radio-frequency continuum emission from evolved stars, the others
being stellar photosphere, chromosphere, and circumstellar dust
\citep{knapp}. Our observations rule out the possibility of a stellar wind
or dust origin for the emission, since the obtained spectra are not
compatible with such hypotheses. The dust would exhibit an
approximate black body spectrum ($S_\nu \propto \nu^2$ in the radio range),
and a stellar wind would have a 0.6 spectral index over a wide range of
frequencies \citep{panagia}.  

We have modelled the observed spectra assuming that the emission arises in
spherical shells around the central stars, with the density decreasing as
$r^{-2}$. The values of electron densities and ionised masses found for
our targets confirm their nature as young PN, since they fall within the
ranges expected for such objects. The data seem to point to a somewhat
steeper than expected slope beyond 5~GHz but, by spectral fitting, we
calculated spectral indexes that match, to within errors, the theoretical
value of -0.1.

IRAS~18062+2410 appears to be optically thick up to 8.4~GHz and
 also shows flux density variations on a time scale of a few years. Such
variations are rarely observed in the post-AGB/young-PN transition phase
and indicate that this star belongs to a group of objects having such
properties as those observed in CRL 618. IRAS 17423-1755 also seems to be
optically thick at high frequency, but the complete spectrum is needed to
confirm this result. Quite interestingly, IRAS 20462+3416, whose overall
centimetre spectrum is very flat, has a larger flux density at 22.4 GHz 
than at lower frequencies. We speculate that this may be due to a
contribution from cold dust.

The high-angular resolution observations that we have presented show that
the ionisation in these targets has already involved the walls of the
cavity around the central star and is not limited to the tenuous gas
within the cavity. All of the targets have been at least partly resolved,
except IRAS 18062+2410. Almost all of the envelopes show two peaks of
emission in a somewhat extended nebulosity that can be explained as an
opacity effect, possibly due to a circumstellar torus, or the onset of
jets. Unlike the other targets, the envelope in IRAS 22023+5249 shows more
of an elongation along an axis than two bright peaks. Such a morphology
would be better explained by the action of jets rather than the presence
of a torus.

\section*{Acknowledgments}

L. Cerrigone acknowledges funding from the Smithsonian Astrophysical
Observatory through the SAO Predoctoral Fellowship programme.  The
National Radio Astronomy Observatory is a facility of the National Science
Foundation operated under cooperative agreement by Associated
Universities, Inc. This research has made use of NASA's Astrophysics Data
System.

\label{lastpage}


\begin{thebibliography}{99}

\bibitem[\protect\citeauthoryear{Aaquist \& Kwok}{1991}]{aaquist91}
Aaquist, O. B. and Kwok, S, 1991, ApJ, 378, 599

\bibitem[\protect\citeauthoryear{Aaquist \& Kwok}{1996}]{aaquist96}
Aaquist, O. B. and Kwok, S, 1996, ApJ, 462, 813

\bibitem[\protect\citeauthoryear{Arkhipova et al.}{2001}]{arkhipova}
Arkhipova, V. P., Ikonnikova, N. P., Noskova, R. I., Kommisarova, G. V., Klochkova, V. G. and Esipov, V. F., 2001, Astr. Letters, 27, 719

\bibitem[\protect\citeauthoryear{Bains et al.}{2003}]{bains}
Bains, I., Bryce, M., Mellema, G.,Redman, M. P. and Thomasson, P., 2003, MNRAS, 340(2), 381

\bibitem[\protect\citeauthoryear{Balick \& Frank}{2002}]{balick&frank}
Balick, B. and Frank, A., 2002, ARA\&A, 40, 439

\bibitem[\protect\citeauthoryear{Blackman et al.}{2001}]{blackman}
Blackman, E. G., Frank, A., Markiel, J. A., Thomas, J. H. and Van Horn, H. M., 2001, Nature, 409, 485

\bibitem[\protect\citeauthoryear{Bogdanov}{2000}]{bogdanov}
Bogdanov, M. B., 2000, Astr. Rep., 44, 685

\bibitem[\protect\citeauthoryear{Conlon et al.}{1993}]{conlon}
Conlon, E. S., Dufton, P. L., McCausland, R. J. H. and Keenan, F. P., 1993, ApJ, 408, 593


\bibitem[\protect\citeauthoryear{Huggins et al.}{2000}]{huggins}
Huggins, P. J., Forveille, T., Bachiller, R. and Cox, P., 2000, ApJ, 544, 889

\bibitem[\protect\citeauthoryear{Kahn \& West}{1985}]{kahn}
Kahn, F. D. and West, K. A., 1985, MNRAS, 212, 837

\bibitem[\protect\citeauthoryear{Karovska et al.}{2007}]{karovska}
Karovska, M., Carilli, C. L., Raymond, J. C. and Mattei, J. A., 2007, ApJ, 661, 1048

\bibitem[\protect\citeauthoryear{Kwok et al.}{1978}]{kwok1978}
Kwok, S., Purton., C. R. and Fitzgerald, P. M., 1978, ApJL, 219, L125

\bibitem[\protect\citeauthoryear{Knapp et al.}{1995}]{knapp}
Knapp, G. R., Bowers, P. F., Young, K. and Phillips, T. G., 1995, ApJ, 455, 293


\bibitem[\protect\citeauthoryear{Latter et al.}{1995}]{latter}
Latter, W.B., Kelly, D.M., Hora, J.L. and Deutsch, L.K., 1995, ApJS, 100, 159




\bibitem[\protect\citeauthoryear{Panagia \& Felli}{1975}]{panagia}
Panagia, A., and Felli, M., 1975, A\&A, 39, 1

\bibitem[\protect\citeauthoryear{Parthasarathy et al.}{2000}]{parthasarathy}
{Parthasarathy}, M. and {Garc{\'{\i}}a-Lario}, P. and {Sivarani}, T. and {Manchado}, A. and {Sanz Fern{\'a}ndez de C{\'o}rdoba}, L., 2000, A\&A, 357, 241

\bibitem[\protect\citeauthoryear{Sahai \& Trauger}{1998}]{sahai98}
Sahai, R. and Trauger, J.T., 1998, AJ, 116(3), 1357


\bibitem[\protect\citeauthoryear{Soker}{2006}]{soker06}
Soker, N., 2006, ApJL, 645, L57


\bibitem[\protect\citeauthoryear{Umana et al.}{2004}]{umana}
Umana, G., Cerrigone, L., Trigilio, C. and Zappal\`a, R.A., 2004, A\&A, 428, 121

\bibitem[\protect\citeauthoryear{Umana et al.}{2008}]{sao}
Umana, G.,  Trigilio, C., Cerrigone, L., Leto, P. and Buemi, C. S., 2008, MNRAS, 386, 1404 


\end{thebibliography}
\end{document}